\documentclass[aps,pra,onecolumn,floatfix,longbibliography,10pt]{revtex4-1}%
\usepackage{graphicx}
\usepackage{dcolumn}
\usepackage{multirow}
\usepackage{bm}
\usepackage{hyperref}
\usepackage{amssymb}
\usepackage{amsmath}
\usepackage{amsfonts}
\usepackage{amssymb}
\usepackage{soul}
\usepackage{ulem}
\usepackage{color}%
\newcommand\inc{{^\textrm{in}}}
\newcommand\res{{^\textrm{res}}}

\begin{document}
\title{Magnetic Feshbach resonances in collisions of $^{23}$Na$^{40}$K with $^{40}$K}
\author{Xin-Yao Wang$^{1,2,4,5,6}$}
\thanks{These authors contributed equally to this work.}
\author{Matthew D. Frye$^{3,*}$}
\email{matthew.frye@durham.ac.uk}
\author{Zhen Su$^{2,4,5}$}
\author{Jin Cao$^{2,4,5}$}
\author{Lan Liu$^{2,4,5}$}
\author{De-Chao Zhang$^{2,4,5}$}
\author{Huan Yang$^{2,4,5}$}
\author{Jeremy M. Hutson$^{3}$}
\email{J.M.Hutson@durham.ac.uk}
\author{Bo Zhao$^{2,4,5}$}
\email{bozhao@ustc.edu.cn}
\author{Chun-Li Bai$^{1,2,6}$}
\email{clbai@cas.cn}
\author{Jian-Wei Pan$^{2,4,5}$}
\email{pan@ustc.edu.cn}
\affiliation{$^{1}$Beijing National Laboratory for Molecular Sciences, Key Laboratory of Molecular Nanostructure and Nanotechnology, CAS Research/Education Center for Excellence in Molecular Sciences, Institute of Chemistry, Chinese Academy of Sciences, Beijing 100190, China}
\affiliation{$^{2}$Hefei National Laboratory for Physical Sciences at the Microscale and Department
of Modern Physics, University of Science and Technology of China,
Hefei, Anhui 230026, China}
\affiliation{$^{3}$Joint Quantum Centre (JQC) Durham-Newcastle, Department of
Chemistry, Durham University, South Road, Durham DH1 3LE, United Kingdom}
\affiliation{$^{4}$Shanghai Branch, CAS Center for Excellence and Synergetic Innovation Center in Quantum
Information and Quantum Physics, University of Science and Technology of China, Shanghai 201315, China}
\affiliation{$^{5}$Shanghai Research Center for Quantum Sciences, Shanghai 201315, China}
\affiliation{$^{6}$University of Chinese Academy of Sciences, Beijing 100049, China}

\begin{abstract}
We present measurements of more than 80 magnetic Feshbach resonances in collisions of
ultracold $^{23}$Na$^{40}$K with $^{40}$K. We assign quantum numbers to a group of low-field
resonances and show that they are probably due to long-range states of the triatomic complex in which
the quantum numbers of the separated atom and molecule are approximately preserved.
The resonant states are not members of chaotic bath of short-range states. Similar resonances are
expected to be a common feature of alkali-metal diatom + atom systems.
\end{abstract}
\date{\today}
\maketitle

\section{Introduction}
Ultracold molecules offer great opportunities to study many areas of physics, including fundamental
chemical reaction dynamics \cite{Balakrishnan:FH2:2001, Tscherbul:JCP:2006,Ospelkaus:react:2010,Rui2017, Hu:2019}, quantum
simulation and computing \cite{DeMille:2002, Ni:Swap:2018, Blackmore:2019, Sawant:qudit:2019}, and
precision measurements of fundamental constants \cite{Safronova:2018, Kobayashi:mep-ratio:2019,
Augenbraun:2020, Fitch:2020}. There has been considerable success in producing ultracold
ground-state molecules, either by associating pairs of ultracold atoms \cite{Ni:KRb:2008,Lang:ground:2008, Danzl:ground:2010,
Takekoshi:RbCs:2014, Molony:RbCs:2014, Wang:2015, Park:NaK:2015, Guo:NaRb:2016,Seesselberg:2018,Yang:K_NaK_resonances:2019,Voges:NaK:2020} or by direct laser
cooling of molecules \cite{Barry:2014, Truppe:MOT:2017, Anderegg:2018, Caldwell:2019, Ding:2020,
Augenbraun:2020}. Understanding collisions of ultracold molecules with themselves and ultracold
atoms is now essential to allow control and further cooling of these molecules \cite{Son:2020,
Valtolina:2020, Matsuda:2020, Jurgilas:2021} to form the dense stable gases needed for many
applications. Further, at such low temperatures, the molecular collisions are highly quantum
mechanical in nature, and are thus of fundamental interest in their own right and a powerful tool
to unravel the nature of molecular interactions.

Magnetic Feshbach resonances are one of the most important features of ultracold collisions
\cite{Chin:RMP:2010}. They occur when the energy of a (quasi-)bound state is tuned across the
incoming collision threshold by an external magnetic field. They have been studied extensively in
atomic scattering, where they are used to tune interaction strengths to control collisional
cooling, study degenerate quantum gases, and investigate strongly interacting systems, amongst
other applications. They are also used to associate pairs of atoms into weakly bound molecules
through magnetoassociation \cite{Kohler:RMP:2006} and provide highly sensitive probes of atomic
interaction potentials.

Magnetic Feshbach resonances in collisions involving molecules are far more challenging to
understand. Full scattering calculations are hampered by the need for huge basis sets and
pathological dependence on uncertainties in the interaction potential \cite{Wallis:LiNH:2011}.
Mayle \emph{et al.}\ \cite{Mayle:2012,Mayle:2013} suggested that the combination of deep
anisotropic potentials and large masses produce a dense forest of chaotic short-range states that
dominate the scattering properties and cause large collisional sticking times. This idea has been
influential, and there is significant experimental evidence for such effects in molecule+molecule
scattering \cite{Gregory:RbCs-collisions:2019, Gregory:RbCs-complex-lifetime:2020, Liu:2020}.
However, the densities of states are now expected to be  much lower than the original predictions
\cite{Christianen:density:2019}, and are also substantially lower for atom+molecule than for
molecule+molecule complexes.
%Therefore, it is not clear whether such a statistical theory will
%apply in the atom+molecule case. Even if it does, it neither explains nor predicts individual
%resonances. A different analysis is needed.
%{\color{red} \st{This raises the question about the nature of the resonant states in atom+molecule collisions. Whether these resonances are caused by the short-range chaotic bound %states which can only be understood in terms of density of states, or they are caused by the long-range bound states, which may be employed to construct a simple collision model, %similar to the atomic Feshbach resonances? Scientific interest on the nature of these resonant states dates back almost one decade ago} \cite{Quemener2012}. \st{Recently, magnetic %Feshbach resonances in atom-molecule collisions have been observed in collisions of $^{23}$Na$^{40}$K with $^{40}$K} \cite{Yang:K_NaK_resonances:2019}. \st{However, the nature of %the resonant states remains elusive since only a few resonances are observed.}}
There is great interest \cite{Quemener2012} in whether atom+molecule scattering is dominated by such chaotic states, or whether the important features are caused by long-range bound states akin to those involved in atom+atom magnetic Feshbach resonances. Recently, magnetic Feshbach resonances have been observed in collisions of $^{23}$Na$^{40}$K with $^{40}$K \cite{Yang:K_NaK_resonances:2019}; however, only a few resonances were observed, so it was not possible to determine the nature of the resonant states or assign quantum numbers to them.

%This is the first time that such resonances have been
%observed in atom-molecule collisions.
 In this paper, we extend the experimental work of Yang \emph{et al.} \cite{Yang:K_NaK_resonances:2019} and carry out a theoretical
study of the resonances. We perform extensive measurements of the Feshbach resonances in 25
collision channels in the magnetic field range between 16 G and 120 G; more than 80 new
atom-molecule Feshbach resonances are observed. We use this large set of results to show that the resonances cluster into groups with clear patterns. We assign both spin and rovibrational character to the bound states responsible for the resonances, and conclude that they are relatively simple long-range states and not chaotic short-range states.

\section{observation of the atom-molecule Feshbach resonances}

The atom-molecule Feshbach resonances are detected by observing the loss of molecules from an
ultracold atom-molecule mixture as a function of magnetic field.
The experimental procedures are
similar to those in \cite{Yang:K_NaK_resonances:2019}.
%Briefly, our experiment starts with about $10^4$ $^{23}$Na$^{40}$K molecules in the rovibrational ground state and about $10^5$ $^{40}$K atoms in a crossed-beam optical dipole trap %at a temperature of about 500 nK.
We first prepare a two-species dark-spot magneto-optical trap of $^{23}$Na and $^{40}$K atoms and load
the atoms into a clover-leaf magnetic trap to perform forced evaporative cooling. The atoms are
then loaded into a crossed-beam optical dipole trap ($\lambda$=1064 nm) to perform further
evaporative cooling. We typically create about $3\times10^{5}$ Na atoms and $1.6\times10^{5}$ K
atoms at a temperature of about 500 nK.

We create weakly bound Feshbach molecules from this atomic mixture by two-photon Raman
photoassociation. The technical details are given in our previous work \cite{Yang:K_NaK_resonances:2019}. We use 3 different target Feshbach states and magnetic fields, each associated with an atomic Feshbach resonance \cite{Park:2012}, depending on the desired
hyperfine level of the molecule. The atoms are prepared in an initial state different from the one
which shows the Feshbach resonance and coupled to the Feshbach state by blue-detuned Raman light
fields. For transitions involving changing the $^{40}$K ($^{23}$Na) state, the single-photon
detuning is about $2\pi\times250$ GHz ($2\pi\times252$ GHz) relative to the D2 transition of
$^{40}$K ($^{23}$Na) atoms, and the Rabi frequency for the corresponding $m_f$-changing atomic
transitions are about $2\pi\times 50$ kHz ($2\pi\times 33$ kHz). Details of the states and resonances
involved are given in Table \ref{Table:states}.

We transfer the molecules from the Feshbach state to a desired hyperfine level of the rovibronic
ground state by stimulated Raman adiabatic passage (STIRAP). The $^{23}$Na$^{40}$K molecule in its rovibrational ground state has 36 hyperfine states due to the
nuclear spins $i_\textrm{Na}=3/2$ and $i_\textrm{K}=4$. In the free molecule these are coupled only
weakly, so at the magnetic field of the experiment they are decoupled with well conserved
projections $m_\textrm{Na}$ and $m_\textrm{K}$, respectively; we therefore label molecular states
$(m_\textrm{Na},m_\textrm{K})_\textrm{mol}$. In this work, 9 different hyperfine levels are prepared. The
hyperfine states $(3/2,-4)_\textrm{mol}$, $(1/2,-4)_\textrm{mol}$, $(-3/2,-3)_\textrm{mol}$,
$(-1/2,-4)_\textrm{mol}$ are created at about 85 G;  the first of these is the absolute ground state. Molecule loss in the atom+molecule mixture is
measured in the magnetic field range between 16 G and 106 G in steps of about 0.5 G. The hyperfine
states $(-1/2,-3)_\textrm{mol}$, $(-1/2,-2)_\textrm{mol}$, $(-3/2,-2)_\textrm{mol}$,
$(-3/2,-1)_\textrm{mol}$, and $(-3/2,-4)_\textrm{mol}$ are prepared at about 102 G. The magnetic
field range studied is between 16 G and 120 G. The intermediate states used in the
STIRAP are the hyperfine levels of a mixed rovibronic state formed from $B^{1}\Pi$
$|v=12,J=1\rangle$ and $c^{3}\Sigma$ $|v=35,J=1\rangle$ \cite{Park:2_photon:2015}. These hyperfine
levels can be labelled by the quantum numbers $F_1=i_{\rm{Na}}+J$ and $F=F_1+i_{\rm{K}}$. The
hyperfine interaction between the nuclear spin of K atoms and the electron spin is small, and the
quantum numbers $F_1$, $m_{F_{1}}$, and $m_F=m_{F_1}+m_{i_{\rm{K}}}$ are approximately good quantum
numbers.  The intermediate states used to prepare different hyperfine levels of the ground states
are listed in Table \ref{Table:states}.

\begin{table*}[tbhp]
\caption{The atomic Feshbach resonances used, the hyperfine levels of the intermediate states, and
the polarization of the STIRAP lasers required to prepare the molecule in the desired hyperfine
level of the rovibronic ground state. The quantum numbers are defined in the main text. The atom-atom resonance at 104 G for $(1,0)_\textrm{Na} +
(9/2,-9/2)_\textrm{K}$ has not been reported before, and its location has been determined by
measuring the binding energy of the Feshbach molecules.
\label{Table:states}}
\begin{ruledtabular}
\begin{tabular}{c|c|c|cccc|cc}
\multirow{2}{*}{Initial  state} & \multirow{2}{*}{Feshbach resonance} & \multirow{2}{*}{\shortstack{Hyperfine level \\ of ground state}} & \multicolumn{4}{c|}{Intermediate state} & \multicolumn{2}{c}{Laser polarization} \\
 & & & $F_1$ & $m_{F_1}$ & $m_{i_\textrm{K}}$ & $m_F$ & Stokes & pump \\ \hline
\multirow{4}{*}{\shortstack{$(1,1)_\textrm{Na}+(9/2,-5/2)_\textrm{K}$}}  & \multirow{4}{*}{\shortstack{$(1,1)_\textrm{Na}+(9/2,-7/2)_\textrm{K}$\\ at 110 G}}
     & $(-1/2,-2)_\textrm{mol}$   & $1/2$ & $-1/2$ & $-2$ & $-5/2$ & $\pi$ & $\pi$ \\
 &  &$(-1/2,-3)_\textrm{mol}$     & $1/2$ & $-1/2$ & $-3$  & $-7/2$ & $\pi$ & $\sigma^{-}$ \\
 &  &$(-3/2,-1)_\textrm{mol}$    & $3/2$ & $-3/2$ & $-1$ & $-5/2$ & $\pi$ & $\pi$ \\
 &  &$(-3/2,-2)_\textrm{mol}$    & $3/2$ & $-3/2$ & $-2$ & $-7/2$ & $\pi$ & $\sigma^{-}$ \\ \hline
\multirow{4}{*}{\shortstack{$(1,1)_\textrm{Na}+(9/2,-7/2)_\textrm{K}$}}  & \multirow{4}{*}{\shortstack{$(1,1)_\textrm{Na}+(9/2,-9/2)_\textrm{K}$\\ at 85 G}}
     & $(3/2,-4)_\textrm{mol}$    & $5/2$ & $3/2$ & $-4$ & $-5/2$ & $\pi$ & $\sigma^{+}$ \\
&  & $(1/2,-4)_\textrm{mol}$     & $5/2$ & $1/2$ & $-4$ & $-7/2$ & $\pi$ & $\pi$ \\
&  & $(-1/2,-4)_\textrm{mol}$    & $1/2$ & $-1/2$ & $-4$ & $-9/2$ & $\pi$ & $\sigma^{-}$ \\
&  & $(-3/2,-3)_\textrm{mol}$    & $3/2$ & $-3/2$ & $-3$ & $-9/2$ & $\pi$ & $\sigma^{-}$ \\ \hline
\multirow{2}{*}{\shortstack{$(1,1)_\textrm{Na}+(9/2,-9/2)_\textrm{K}$}}  & \multirow{2}{*}{\shortstack{$(1,0)_\textrm{Na}+(9/2,-9/2)_\textrm{K}$\\ at 104 G}}
     & $(-3/2,-4)_\textrm{mol}$   & $3/2$ & $-3/2$ & $-4$ & $-11/2$ & $\pi$ & $\sigma^{-}$ \\ & & & & & & & &
\end{tabular}
\end{ruledtabular}
\end{table*}

In the experiment, the pump laser coupling the Feshbach state and the
intermediate state is a 805 nm TA-Pro laser, and the Stokes laser coupling the ground state and
the intermediate state is a 567 nm TA-SHG laser. The two lasers are locked to a ULE cavity to
suppress the noises. The Stokes light is
$\pi$ polarized and the required polarization of the pump light is $\pi$, $\sigma^{+}$ or
$\sigma^{-}$. Due to the limited optical access, the two lasers propagate perpendicular to the
magnetic field. The $\pi$ polarization is achieved by setting the polarization of the lasers
parallel to the magnetic field. The $\sigma$ polarization is applied by setting the polarization of
the pump laser to $\sigma^{+}+\sigma^{-}$ linear polarization. A detailed theoretical analysis of
the coupling strength between the Feshbach states and the hyperfine levels of the intermediate
states is given in Ref.\ \onlinecite{LiuY2020}. There are many possible pathways for STIRAP
and in principle many hyperfine levels of the rovibronic ground state can be
populated. We have chosen hyperfine levels of the excited states so that the desired pathway
dominates over the others and the target hyperfine level is
prepared. The purity of the hyperfine states of the ground-state molecule is checked by observing
that the molecule number after a round-trip STIRAP does not oscillate as a function of the hold
time \cite{Liu:2019}.

 The $^{40}$K atom has electron spin $s=1/2$ in addition to nuclear spin $i_\textrm{K}=4$; these
couple strongly to form $f=9/2$ and $7/2$ at zero field, with projection $m_f$. At the fields of
the experiment, $f$ is a nearly good quantum number and we label the atomic states of $^{40}$K by $(f, m_f
)_\textrm{at}$. The atoms are prepared in the states $(9/2,-9/2)_\textrm{at}$,
$(9/2,-7/2)_\textrm{at}$, and $(9/2,-5/2)_\textrm{at}$.

After the atom-molecule mixture is prepared, we ramp the magnetic field to the desired value in a
few milliseconds and then hold the atom-molecule mixture for about 7 ms. We program the profile of
current of the coil to compensate the residual magnetic field caused by the eddy currents, so that
the magnetic field can reach the desired value in about 4 to 6 ms. During the hold time, the
magnetic field drifts and fluctuates, and the magnitude of the uncertainty of the magnetic field is
proportional to the magnetic field range we ramp. When the magnetic field is ramped from about 102
G to about 16 G, the uncertainty of the magnetic field during the hold time is about 800 mG. The
hold time of the atom-molecule mixture is controlled by removing the $^{40}$K atoms using resonant
light pulses. After that, the magnetic field is ramped back to the initial value in 3 ms. After
about 20 ms when the magnetic field is stable, we transfer the ground-state molecules to the
Feshbach state by reverse STIRAP. The Feshbach molecules are either directly detected by the
standard time-of-flight absorption imaging or are dissociated into free atoms, which are then
detected by absorption imaging. During the hold time, molecules are lost from the trap due
to inelastic 2-body atom-molecule collisions or 3-body recombination. At resonance, both these loss
processes are enhanced and we see characteristic dips in the remaining molecule number as a
function of field. We identify these loss features as the atom-molecule Feshbach resonances.
%The Feshbach resonances appear as characteristic dips in the remaining number of molecules as a
%function of field.
Narrow resonances in the low-field region are challenging to detect in the current experiment, both
because the long field ramp to the target field creates uncertainties in the final field, and
because the signal-to-noise ratio is reduced if molecules are lost in traversing higher-field resonances on
the way.

We have searched for Feshbach resonances in 25 of the 27 possible combinations of these 9 molecular
states and 3 atomic states. For the combinations $(-3/2,-4)_\textrm{mol} + (9/2,-5/2)_\textrm{at}$
and $(-1/2,-3)_\textrm{mol} + (9/2,-5/2)_\textrm{at}$, the background loss rates are very large and
our method is not applicable. The resonances for $(-3/2,-2)_\textrm{mol}$ at magnetic fields
between 43 G and 120 G were studied in our previous work \cite{Yang:K_NaK_resonances:2019}, but
here we extend the magnetic field range down to 16 G. In total, we observe 84 new atom-molecule
Feshbach resonances. Example loss features are shown in Fig.\ \ref{fig:sam}. The other 80 loss features are displayed in the supplementary material.
%The loss features are fitted to
%Gaussian functions and the resulting resonance positions $B_0$ and widths $\delta$ are summarized in Tables II and III of Supplemental Material \cite{Supp_KNaK}.
We fit the loss features with Gaussian functions to extract a position $B_0$ and width $\delta$.
Note that the observed width of the loss feature $\delta$ is not the same quantity as the theoretical width
$\Delta$ that describes the pole in scattering length associated with the resonance. We give
results for all resonances observed in Tables \ref{Table:low} and \ref{Table:high}, for the low-
and high-field groups of resonances, respectively.

\begin{figure}[tb]
\includegraphics[width=12cm]{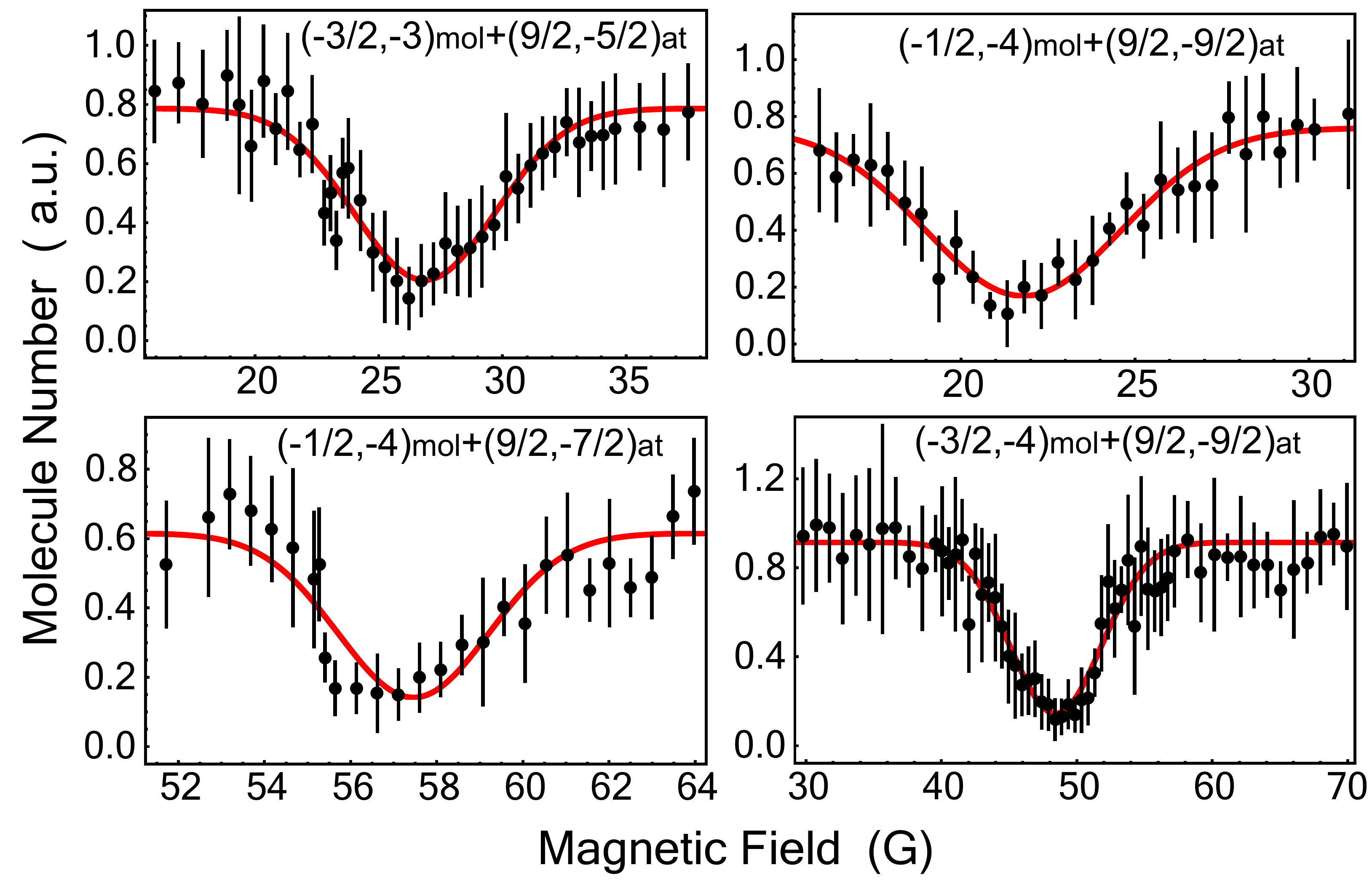}
\caption{\label{fig:sam} Example loss features for different collision channels. The remaining molecule numbers are shown as a function of the magnetic field. The solid lines are the fitted Gaussian functions. Error bars represent the standard deviation.}
\end{figure}

\section{Assignment of the low-field atom-molecule Feshbach resonances}

In order to interpret these resonances, we need to describe the resonant bound states that cause
them and assign their quantum numbers $X\res$. We look for explanations in terms of resonances of
as few different characters as possible;  here the character is defined by the rovibrational state
of the triatomic complex and the changes $\Delta X = X\res - X\inc$ between incoming and resonant
states for the remaining quantum numbers.
If a resonance of a particular character exists at one
threshold, a resonance of that character  is expected to exist at every threshold where it is allowed.

\begin{table*}
\caption{Atom-molecule Feshbach resonances in the low-field group, along with the assigned quantum numbers for the bound states that cause them as described in the main text. \label{Table:low}}
\centering
\begin{ruledtabular}
\begin{tabular}{cc|cc|ccccc}
\multicolumn{2}{ c |}{Initial State}		& \multicolumn{2}{ c |}{Measured}	& \multicolumn{5}{ c }{Assigned quantum numbers}					\\
$m_\textrm{Na}$ 	& $m_\textrm{K}$	& $B_0$ (G)	& $\delta$ (G)		& $m_f$	& $m_\textrm{Na}$	& $m_\textrm{K}$	& $N$	& $M_N$	\\ \hline
\multicolumn{9}{c}{ Initial state of the K atom $(f,m_f)=(9/2,-9/2)$} \\ \hline
$-3/2$			& $-3$			& 20.9		& 3.2				& $-7/2$	& $-3/2$			& $-4$			 & 1 or 2		& 0		\\
$-3/2$			& $-2$			& 19.4		& 2.6				& $-7/2$	& $-3/2$			& $-3$			 & 1 or 2		& 0		\\
$-3/2$			& $-1$			& 19.0		& 1.0				& $-7/2$	& $-3/2$			& $-2$			 & 1 or 2		& 0		\\
$-1/2$			& $-4$			& 21.8		& 3.9				& $-7/2$	& $-3/2$			& $-4$			 & 1 or 2		& 0		\\
$-1/2$			& $-3$			& 19.3		& 5.4				& $-7/2$	& $-3/2$ or $-1/2$	& $-3$ or $-4$		& 1 or 2		& 0		\\
$-1/2$			& $-2$			& 19.3		& 2.9				& $-7/2$	& $-3/2$ or $-1/2$	& $-2$ or $-3$		& 1 or 2		& 0		\\
$1/2$			& $-4$			& 16.6		& 0.7				& $-7/2$	& $1/2$			& $-4$			 & 1 or 2		& $-1$	\\
$1/2$			& $-4$			& 20.9		& 3.9				& $-7/2$	& $-1/2$			& $-4$			 & 1 or 2		& 0		\\
$3/2$			& $-4$			& 17.1		& 0.9				& $-7/2$	& $3/2$			& $-4$			 & 1 or 2		& $-1$	\\
$3/2$			& $-4$			& 20.1		& 1.7				& $-7/2$	& $1/2$			& $-4$			 & 1 or 2		& 0		\\ \hline
\multicolumn{9}{c}{Initial state of the K atom $(f,m_f)=(9/2,-7/2)$} \\ \hline
$-3/2$			& $-4$			& 19.6		& 2.0				& $-5/2$	& $-3/2$			& $-4$			 & 1 or 2		& $-1$	\\
$-3/2$			& $-3$			& 22.4		& 2.8				& $-5/2$	& $-3/2$			& $-4$			 & 1 or 2		& 0		\\
$-3/2$			& $-2$			& 23.6		& 3.6				& $-5/2$	& $-3/2$			& $-3$			 & 1 or 2		& 0		\\
$-3/2$			& $-1$			& 24.3		& 2.5				& $-5/2$	& $-3/2$			& $-2$			 & 1 or 2		& 0		\\
$-1/2$			& $-4$			& 23.5		& 5.2				& $-5/2$	& $-3/2$			& $-4$			 & 1 or 2		& 0		\\
$-1/2$			& $-3$			& 22.9		& 5.9				& $-5/2$	& $-3/2$ or $-1/2$	& $-3$ or $-4$		& 1 or 2		& 0		\\
$-1/2$			& $-2$			& 23.1		& 4.0				& $-5/2$	& $-3/2$ or $-1/2$	& $-2$ or $-3$		& 1 or 2		& 0		\\
$1/2$			& $-4$			& 24.8		& 5.7				& $-5/2$	& $-1/2$			& $-4$			 & 1 or 2		& 0		\\
$3/2$			& $-4$			& 19.8		& 0.6				& $-5/2$	& $3/2$			& $-4$			 & 1 or 2		& $-1$	\\
$3/2$			& $-4$			& 24.9		& 1.5				& $-5/2$	& $1/2$			& $-4$			 & 1 or 2		& 0		\\ \hline
\multicolumn{9}{c}{Initial state of the K atom $(f,m_f)=(9/2,-5/2)$} \\ \hline
$-3/2$			& $-3$			& 26.8		& 3.9				& $-3/2$	& $-3/2$			& $-4$			 & 1 or 2		& 0		\\
$-3/2$			& $-2$			& 29.7		& 5.2				& $-3/2$	& $-3/2$			& $-3$			 & 1 or 2		& 0		\\
$-3/2$			& $-1$			& 31.7		& 1.9				& $-3/2$	& $-3/2$			& $-2$			 & 1 or 2		& 0		\\
$-1/2$			& $-4$			& 26.1		& 4.9				& $-3/2$	& $-3/2$			& $-4$			 & 1 or 2		& 0		\\
$-1/2$			& $-2$			& 27.2		& 1.0				& $-3/2$	& $-3/2$ or $-1/2$	& $-2$ or $-3$		& 1 or 2		& 0		\\
$1/2$			& $-4$			& 25.1		& 0.4				& $-3/2$	& $1/2$			& $-4$			 & 1 or 2		& $-1$	\\
$1/2$			& $-4$			& 29.7		& 3.0				& $-3/2$	& $-1/2$			& $-4$			 & 1 or 2		& 0		\\
$3/2$			& $-4$			& 31.2		& 3.1				& $-3/2$	& $1/2$			& $-4$			 & 1 or 2		& 0		\\
\end{tabular}
\end{ruledtabular}
\end{table*}

In the three-atom complex, the diatom rotation $n$ couples to the angular momentum $L$ of the atom
and diatom around one another to form $N$, which represents the total spin-free angular momentum.
The quantum numbers $(v,n,L,N)$, with $v$ the diatom vibrational quantum number, provide a good representation in the long-range region where the
interaction potential is weak, except for coupling to spins and the magnetic field. The incoming
states relevant to the experiment all have $v\inc=n\inc=0$; since $L\inc=0$ dominates at the temperature of the experiment, this implies $N\inc=0$. The electronic interaction
potential is strong and highly anisotropic at short range, which results in complicated states with
strongly mixed $n$, $L$, and $v$, but it does not mix $N$.

\begin{table*}
\caption{Atom-molecule Feshbach resonances in the high-field group, along with the assigned quantum numbers for the bound states that cause them as described in the main text. \label{Table:high}}
\centering
\begin{ruledtabular}\footnotesize
\begin{tabular}{cc|cc|c}
\multicolumn{2}{ c |}{Initial State}		& \multicolumn{2}{ c |}{Measured}	& \multicolumn{1}{ c }{Assigned quantum number}					\\
$m_\textrm{Na}$ 	& $m_\textrm{K}$	& $B_0$ (G)	& $\delta$ (G)		& $m_f$	\\ \hline
\multicolumn{5}{c}{Initial state of the K atom $(f,m_f)=(9/2,-9/2)$} \\ \hline
$-3/2$			& $-4$			& 48.4		& 4.8				& $-7/2$	\\
$-3/2$			& $-3$			& 47.8		& 3.0				& $-7/2$	\\
$-3/2$			& $-2$			& 48.5		& 2.1				& $-7/2$	\\
$-3/2$			& $-2$			& 90.3		& 2.2				& $-7/2$	\\
$-3/2$			& $-1$			& 50.6		& 1.1				& $-7/2$	\\
$-3/2$			& $-1$			& 59.6		& 1.3				& $-7/2$	\\
$-3/2$			& $-1$			& 95.3		& 1.9				& $-7/2$	\\
$-1/2$			& $-4$			& 43.4		& 0.6				& $-7/2$	\\
$-1/2$			& $-4$			& 47.7		& 1.9				& $-7/2$	\\
$-1/2$			& $-3$			& 45.1		& 0.5				& $-7/2$	\\
$-1/2$			& $-3$			& 48.8		& 1.7				& $-7/2$	\\
$-1/2$			& $-2$			& 50.5		& 1.3				& $-7/2$	\\
$-1/2$			& $-2$			& 59.7		& 3.4				& $-7/2$	\\
$-1/2$			& $-2$			& 94.9		& 0.6				& $-7/2$	\\
$1/2$			& $-4$			& 43.4		& 0.4				& $-7/2$	\\
$1/2$			& $-4$			& 45.4		& 0.4				& $-7/2$	\\
$1/2$			& $-4$			& 48.8		& 1.1				& $-7/2$	\\
$3/2$			& $-4$			& 81.5		& 0.3				& $-7/2$	\\
$3/2$			& $-4$			& 83.4		& 0.3				& $-7/2$	\\ \hline
\multicolumn{5}{c}{Initial state of the K atom $(f,m_f)=(9/2,-7/2)$} \\ \hline
$-3/2$			& $-4$			& 63.2		& 12.3			& $-5/2$	\\
$-3/2$			& $-3$			& 52.0		& 0.5				& $-7/2$	\\
$-3/2$			& $-3$			& 57.6		& 5.3				& $-7/2$	\\
$-3/2$			& $-3$			& 102.9		& 3.0				& $-7/2$	\\
$-3/2$			& $-2$			& 54.6		& 0.6				& $-7/2$	\\
$-3/2$			& $-2$			& 59.4		& 3.1				& $-5/2$	\\
$-3/2$			& $-2$			& 101.1		& 0.6				& $-7/2$	\\
$-3/2$			& $-2$			& 106.3		& 1.7				& $-7/2$	\\
$-3/2$			& $-1$			& 57.7		& 0.7				& $-5/2$	\\
$-3/2$			& $-1$			& 62.2		& 2.2				& $-5/2$	\\
$-3/2$			& $-1$			& 106.0		& 0.7				& $-7/2$	\\
$-1/2$			& $-4$			& 57.5		& 2.5				& $-5/2$	\\
$-1/2$			& $-3$			& 59.4		& 2.7				& $-5/2$	\\
$-1/2$			& $-3$			& 68.7		& 3.7				& $-5/2$	\\
$-1/2$			& $-3$			& 101.0		& 0.3				& $-7/2$	\\
$-1/2$			& $-3$			& 106.5		& 1.1				& $-7/2$	\\
$-1/2$			& $-2$			& 62.3		& 2.2				& $-5/2$	\\
$-1/2$			& $-2$			& 72.1		& 1.8				& $-5/2$	\\
$1/2$			& $-4$			& 51.8		& 0.3				& $-7/2$	\\
$1/2$			& $-4$			& 59.2		& 0.9				& $-5/2$	\\ \hline
\multicolumn{5}{c}{Initial state of the K atom $(f,m_f)=(9/2,-5/2)$} \\ \hline
$-3/2$			& $-3$			& 57.8		& 3.6				& $-7/2$	\\
$-3/2$			& $-3$			& 63.1		& 1.2				& $-5/2$	\\
$-3/2$			& $-3$			& 70.0		& 2.4				& $-5/2$	\\
$-3/2$			& $-3$			& 102.4		& 2.0				& $-7/2$	\\
$-3/2$			& $-2$			& 68.2		& 1.0				& $-5/2$	\\
$-3/2$			& $-2$			& 74.5		& 3.7				& $-5/2$	\\
$-3/2$			& $-2$			& 83.5		& 1.6				& $-3/2$	\\
$-3/2$			& $-1$			& 72.9		& 1.2				& $-5/2$	\\
$-3/2$			& $-1$			& 79.3		& 2.3				& $-3/2$	\\
$-3/2$			& $-1$			& 87.2		& 0.7				& $-3/2$	\\
$-3/2$			& $-1$			& 90.1		& 1.2				& $-3/2$	\\
$-1/2$			& $-4$			& 56.8		& 4.0				& $-7/2$	\\
$-1/2$			& $-4$			& 62.6		& 1.0				& $-5/2$	\\
$-1/2$			& $-4$			& 69.5		& 1.3				& $-5/2$	\\
$-1/2$			& $-4$			& 78.8		& 2.0				& $-3/2$	\\
$-1/2$			& $-4$			& 100.9		& 2.1				& $-7/2$	\\
$-1/2$			& $-2$			& 55.9		& 2.1				& $-7/2$	\\
$-1/2$			& $-2$			& 79.5		& 2.3				& $-3/2$	\\
$-1/2$			& $-2$			& 83.3		& 0.8				& $-3/2$	\\
$-1/2$			& $-2$			& 90.4		& 1.9				& $-3/2$	\\
$1/2$			& $-4$			& 51.7		& 3.2				& $-7/2$	\\
$1/2$			& $-4$			& 63.7		& 0.3				& $-5/2$	\\
$1/2$			& $-4$			& 73.7		& 0.8				& $-5/2$	\\
$1/2$			& $-4$			& 77.7		& 0.4				& $-3/2$	\\
$1/2$			& $-4$			& 83.7		& 1.2				& $-3/2$	\\
$3/2$			& $-4$			& 50.5		& 0.3				& $-7/2$	\\
$3/2$			& $-4$			& 56.2		& 1.4				& $-7/2$	\\
\end{tabular}
\end{ruledtabular}
\end{table*}

The interaction potential between ground-state NaK and K is strongly anisotropic, but
there is only a single Born-Oppenheimer
potential-energy surface that correlates with the incoming states. There is thus no strong coupling
due to differences between two potentials, as there is in collisions of two alkali-metal atoms, or
as there would be in collisions of an electron-spin-doublet molecule with an alkali-metal atom. The important
couplings are expected to include various Zeeman, spin-rotation, and hyperfine couplings
\cite{Frye:triatomic-complexes:2021, Frye:triatomic-threshold-complexes:2021}.
A detailed description and estimation of the strength of
these couplings is beyond the scope of this paper, but they are expected to be fairly weak and to
act essentially perturbatively. This means that the bound states can be usefully labeled with the
same spin quantum numbers as the incoming states, even if there is significant coupling of $s$ to the molecular nuclear spins at short range. This situation has some similarities with that in
alkali+$^1$S systems \cite{Zuchowski:RbSr:2010,Brue:LiYb:2012,Brue:AlkYb:2013,Barbe:RbSr:2018,Yang:CsYb:2019}.

We consider first the group of resonances between 16 and 32 G, which we refer to as the low-field
group. 28 such resonances are currently observed. The general layout of the resonances in
field must come from the atomic quantum numbers $(f, m_f)$, since the associated magnetic moments
are far larger than the magnetic moments of the molecule; one or both of $f$ and $m_f$ must change
for the resonant state to cross the incoming threshold. The lowest state of $^{40}$K is
$(9/2,-9/2)_\textrm{at}$ and is fully spin-stretched, so the resonances observed for this incoming
state must have $\Delta m_f=+1$. This selection rule can explain resonances for all three atomic states investigated, so it seems likely that this character applies to all resonances in the low-field group.

\begin{figure}[tb]
\includegraphics[width=12cm]{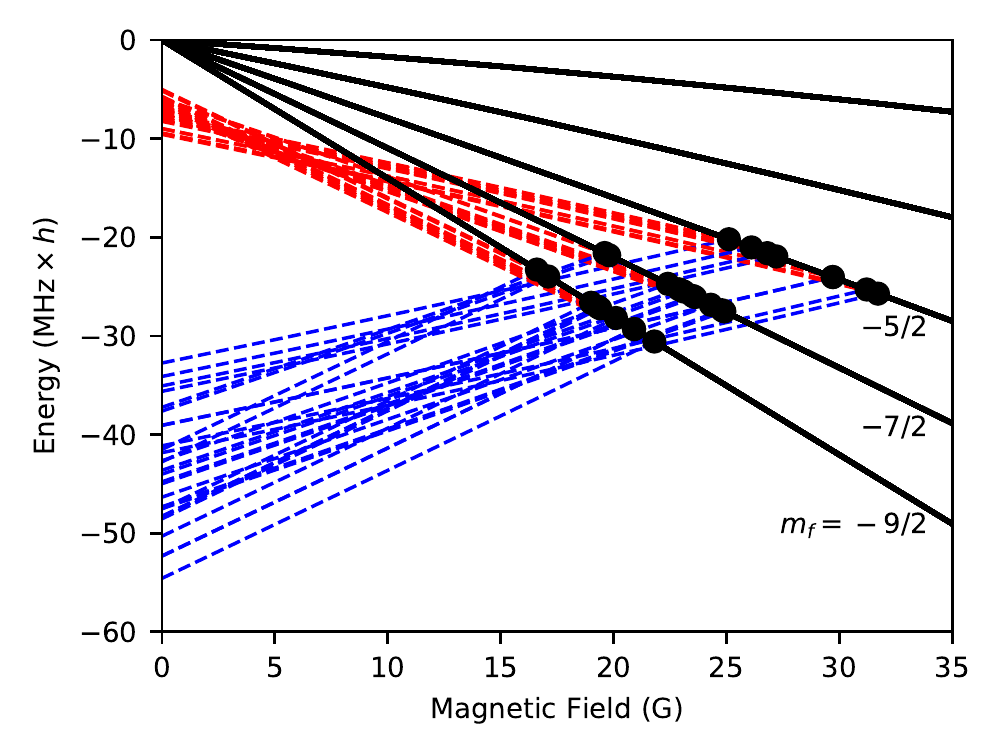}
\caption{\label{fig:low} Observed positions of resonances in the low-field group (black dots), with
projections back to zero field assuming $f\res=9/2, \Delta m_f=+1$ (red dashed lines) and
$f\res=7/2, \Delta m_f=+1$ (blue dashed lines). Black solid lines indicate thresholds. The weak
Zeeman effect of the molecular spins is neglected.}
\end{figure}

In order to determine $f\res$ for these resonances, we consider what each possibility would imply
about the levels. We expect that, towards zero field, sublevels with different $m_f\res$ will
approach one another to form a single clear group, which may have internal structure due to
hyperfine couplings in the triatomic complex. We therefore calculate the
implied energy of the resonant state, determining $m_f\res$ from $\Delta m_f=+1$ and considering
$f\res=9/2$ or $7/2$. We project each bound state back to zero field, assuming it remains at a
constant binding energy below its corresponding threshold; this neglects mixing between different
levels in the group, but should be a good first approximation. The results of this are shown in
Fig.\ \ref{fig:low}.

The red lines in Fig.\ \ref{fig:low} show that the pattern of resonances is consistent with a closely spaced group of zero-field bound states with $f\res=9/2$. These have binding energies centred near 8 MHz and spread across about 5 MHz. Such a closely spaced group is unlikely to occur by coincidence. They produce a tightly bunched group of resonances at each atomic threshold, spread over less than 6~G for each $m_f$. Conversely, the blue lines for an assignment to $f\res=7/2$ diverge towards zero field and would imply zero-field binding energies spread across at least 20 MHz.  However, in this case the states with a given value of $m_f\res$ would be expected to cause additional resonances with $\Delta m_f=0$ and $-1$ when they cross the higher thresholds. In fact the number of resonances at higher thresholds is similar to that at the $m_f=-9/2$ threshold. It is thus probable that the low-field resonances arise from states with $f\res=9/2$.

To assign any further quantum numbers, we must look at the detailed patterns of resonances. The
low-field resonances are sometimes observed in pairs at the same threshold, so there must be
(at least) two characters of resonances in this group. For the thresholds where both resonances are
observed, there is a consistent pattern that the one at higher field is wider than the one at lower
field, and we label these as having characters A and B respectively. For thresholds where there is
only one resonance observed, it is possible that one of the resonances is systematically missing
due to selection rules, but it is also possible that the second resonance may be present but not
observed.

There is one resonance observed for the incoming combination $(-3/2,-4)_\textrm{mol} +
(9/2,-7/2)_\textrm{at}$, for which the molecule is in its spin-stretched state. This resonance must
have $\Delta m_\textrm{mol}=\Delta m_\textrm{Na}+\Delta m_\textrm{K}\geq 0$. Combined with $\Delta
m_f=+1$, as discussed above, and conservation of total projection, $\Delta M_\textrm{tot}=\Delta
m_\textrm{mol}+\Delta m_f+\Delta M_N=0$, this implies that $\Delta M_N \leq -1$ and so $N\res \neq
0$. States with different $N\res$ are expected to be far  apart in energy, so all states in this
group must have the same value of $N\res$, but may have different $M_N\res$.
%We will proceed
%on the supposition that $N\res =1$, although we cannot rule out higher values. This suggests that
%the two characters of resonances in the low-field group are $(\Delta M_N,\Delta m_\textrm{mol})
%=(-1,0)$ and $(0,-1)$.
There are no terms in the fine, hyperfine or Zeeman Hamiltonian that can directly change $N$ by more than 2, and only the small nuclear electric quadrupole interaction has $\Delta M_N>1$ \cite{Frye:triatomic-threshold-complexes:2021}; the most likely interpretation is therefore that $N\res =1$ or 2, and that the two characters of resonances in the low-field group are $(\Delta M_N,\Delta m_\textrm{mol}) =(-1,0)$ and $(0,-1)$. The latter is missing at  the threshold $(-3/2,-4)_\textrm{mol} + (9/2,-7/2)_\textrm{at}$ because the molecule is
spin-stretched, so the observed resonance must have character $(-1,0)$. However, it is not easily
assigned as either A or B from its position and width.

For the incoming combination $(-3/2,-4)_\textrm{mol} + (9/2,-9/2)_\textrm{at}$, no resonance is
observed at all, but we expect one with character $(-1,0)$. This resonance might be missed either
because it is outside the region of observation or because it is too narrow. Both of these are more
likely if it is of character B, so we tentatively assign character A to be $(0,-1)$ and character B
to be $(-1,0)$.

This assignment determines the spin character of the resonant bound states almost completely. The
resulting quantum numbers are included in Tables II and III. The only remaining ambiguities are in cases where different values of $\Delta
m_\textrm{Na}$ and $\Delta m_\textrm{K}$ can sum to give the same $\Delta m_\textrm{mol}$. This
might, for example, lead to additional resonances for collisions involving
$(-1/2,-3)_\textrm{mol}$, but the present experiments have not resolved any such features.

Our assignment implies that there is a rovibrational state of the triatomic complex with
$N\res\geq1$ that lies very close to threshold. If this has $f\res=9/2$, it is bound by less than 10~MHz.  For a
system with unknown scattering length, an argument based on single-channel quantum defect theory \cite{Gao:2000, Frye:triatomic-threshold-complexes:2021} indicates there is a
9\% (4\%) prior probability of finding a p (d)-wave state within 10 MHz of threshold.
The real system is more complex than a
single channel, but this estimate is appropriate for a state so close to threshold.
Conversely, if the state has $f\res=7/2$, its binding energy must be in a narrow range within 50 MHz of the atomic hyperfine splitting; the corresponding probability for this is 2\% for either a p-wave or a d-wave state.
The final
possibility is a state with $n>0$, for which $N>0$ is allowed even if $L=0$. The probability of
such a short-range bound state existing in the top 10~MHz of the well is low for any individual
value of $n$ and $v$, but much larger when the full phase space is considered
\cite{Christianen:density:2019}. We estimate the density of states near threshold to be 0.3
GHz$^{-1}$ for NaK+K \cite{Frye:triatomic-complexes:2021}, so there is only about a 0.3\% prior
probability of finding such a state within 10 MHz of threshold.

The bound state that causes the low-field resonances is therefore most likely a very weakly bound
long-range state. Such a bound state will have the character predominantly of the separated atom
and molecule in their respective states. This presents a physical picture where the complicated
short-range rovibrational couplings are relatively unimportant and the relevant states of the
complex are remarkably simple. This is in sharp contrast to the picture of resonances caused by
chaotic short-range states, as suggested by Mayle \emph{et al.} \cite{Mayle:2012}.

The long-range nature of the states responsible for the low-field resonances suggests that similar
resonances will be present in other alkali-metal atom+diatom systems. For each value of $L\res$, there will always be
one long-range rovibrational state of the triatomic complex within a certain ``bin" of energy below
threshold \cite{Gao:2000}. For low $L\res$, such a state is likely to couple to the incoming states
to produce a group of magnetically tunable Feshbach resonances below a certain characteristic
magnetic field. The depth of the top bin and the corresponding characteristic field depend on the
masses of the atom and molecule and on the long-range dispersion coefficient.
If $f\res=9/2$, the least-bound state for $^{23}$Na$^{40}$K+$^{40}$K lies in the upper part of its bin, but for other systems this may not be the case; nevertheless, analogous resonances are typically expected at fields below 400 G for heavier systems
\cite{Frye:triatomic-threshold-complexes:2021}.

The remaining 66 resonances span from 43 G to 107 G. We will refer to these as the high-field
group. They do not form so coherent a pattern as the low-field group, and may arise from sets of
resonances with several different characters. In the same way as for the low-field group, the resonances at
thresholds with $m_f\inc=-9/2$ must have $\Delta m_f=+1$. However, at thresholds with
$m_f\inc=-7/2$ there are a larger number of resonances than for $m_f\inc=-9/2$. This increase is
probably caused by resonances with $\Delta m_f=0$, which are allowed at thresholds with
$m_f\inc\ge-7/2$. The further increase in the number of resonances at thresholds with
$m_f\inc=-5/2$ can similarly be attributed to $\Delta m_f=-1$. As discussed above, at least one of
$f$ or $m_f$ must change between the incoming and bound states, so $\Delta m_f=0$ implies $\Delta f
\neq 0$. We thus conclude that the high-field resonances probably arise from resonant states with
$f\res=7/2$.

\begin{figure}[tb]
\includegraphics[width=12cm]{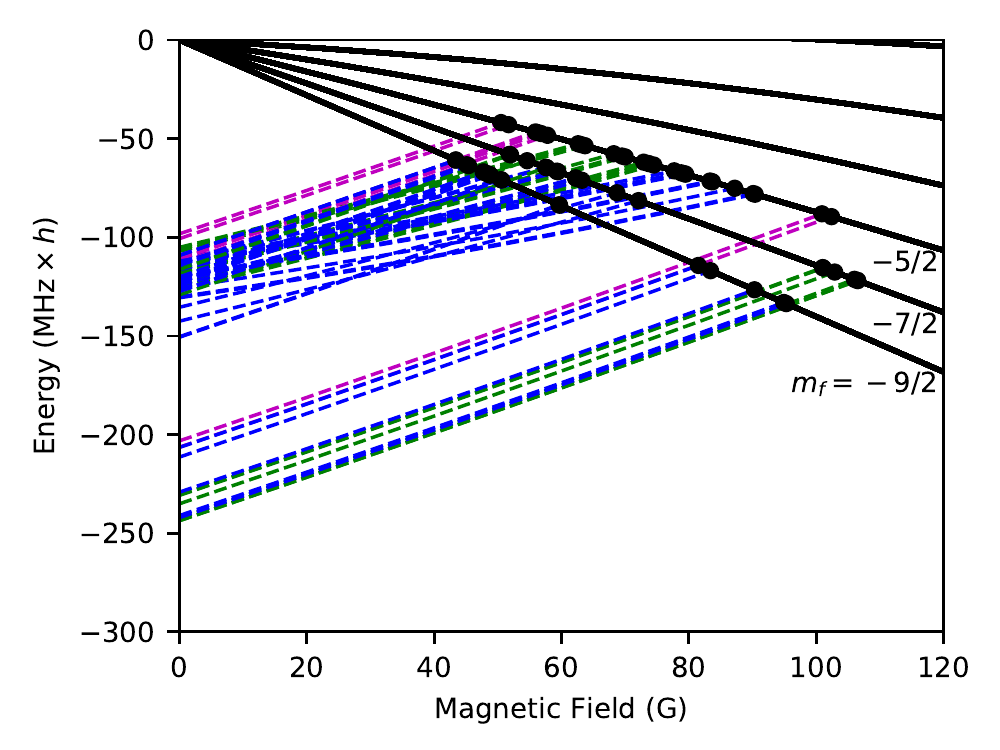}
\caption{\label{fig:high} Observed positions of resonances in the high-field group (black dots),
with projections back to zero field assuming $f\res=7/2, \Delta m_f=+1$ (blue dashed lines);
$f\res=7/2, \Delta m_f=0$ (green dashed lines); or $f\res=7/2, \Delta m_f=-1$ (magenta dashed
lines). Each resonance is tentatively assigned to one of these options. Black solid lines indicate
thresholds. The weak Zeeman effect of the molecular spins is neglected.}
\end{figure}

We project the bound states for this group of resonances back to zero field as described above. We
take $f\res=7/2$ for all resonances in this group, but $m_f\res$ is not unambiguously assigned for
$m_f\inc\ge-7/2$, so we pick values that produce as simple a pattern as possible. The implied
bound-state pattern is shown in Fig.\ \ref{fig:high}. Many of the states appear to converge towards
a zero-field grouping between $-105$ and $-130$ MHz, relative to the $f\inc=9/2$ thresholds, with a
few surrounding outliers. This is not intended to provide a definitive assignment for individual
states, but to indicate that a plausible interpretation along these lines may be possible. The
zero-field states between $-200$ and $-250$ MHz seem to form a separate group, but there are too
few of them to identify clear patterns.

The character of the states responsible for the high-field resonances remains uncertain, but there
are a few likely possibilities. They might arise from states with $v\res=n\res=0$, bound by just
the 1.3 GHz of the atomic hyperfine splitting; they might arise from a low-lying excited rotational state
with $n\res>0$, bound by the rotational excitation energy $bn(n+1)$  where $b/h=2.9$ GHz, either with or without atomic hyperfine excitation; or they might arise from a complicated short-range state of the kind suggested
by Mayle \emph{et al.} \cite{Mayle:2012}. The first of these possibilities would still have
substantial long-range character, with decreasing amounts for the subsequent possibilities.

\section{conclusion}

In summary, we have performed a detailed experimental and theoretical study of the ultracold
atom-molecule Feshbach resonances between $^{23}$Na$^{40}$K and $^{40}$K. More than 80 new Feshbach
resonances have been observed for a range of initial states. We have presented a detailed
theoretical interpretation of the low-field group of resonances and assigned quantum numbers to the
bound states that cause them.  This allows us to determine the nature of the resonant states:  they are likely to be remarkably simple long-range states with the
character of the separated atom and molecule. Such states are very different from the chaotic
short-range states proposed by Mayle \emph{et al.} \cite{Mayle:2012}. The simplicity of the
long-range states allows understanding and offers the possibility of controlling collisions using
these resonances. We expect similar resonances to be observable in other alkali-metal atom + diatom
systems.

\bigskip

\begin{acknowledgments}
This work was supported by the National  Key R\&D Program of China (under Grant No.
2018YFA0306502), the National Natural Science Foundation of China (under Grant No. 11521063, 11904355),  the
Chinese Academy of Sciences, the Anhui Initiative in Quantum Information Technologies, Shanghai Municipal Science and Technology Major Project (Grant No.2019SHZDZX01), Shanghai Rising-Star Program (Grant No. 20QA1410000). This work was supported by the U.K. Engineering and Physical Sciences Research Council (EPSRC) Grant No.\
EP/P01058X/1.

X.-Y. Wang led the experimental work, and Matthew D. Frye led the theoretical work.
\end{acknowledgments}

\section*{Data availability statement}
All data that support the findings of this study are included within the article (and any supplementary
information files).

\thanks{These authors contributed equally to this work.}

\newpage

\section*{supplementary materials}

In the supplementary material, the loss spectrum for the other 80 resonances between $^{23}$Na$^{40}$K and $^{40}$K are shown in the Figs. S1-S6.

\begin{figure}[h]
\renewcommand\thefigure{S1}
\includegraphics[width=16cm]{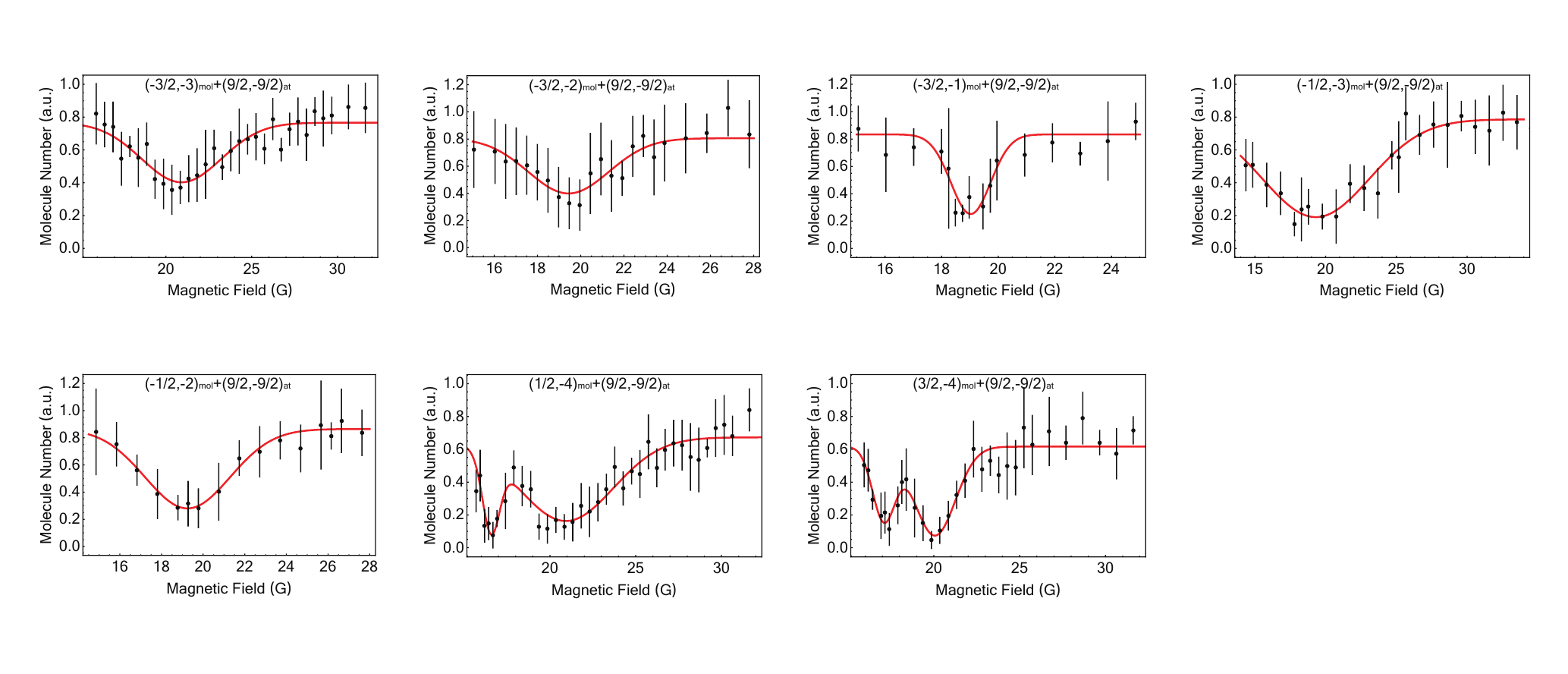}
\caption{\label{fig:sam} The low-field resonances with the initial state of the $^{40}$K atom in $(9/2,-9/2)$. The remaining molecule numbers are shown as a function of the magnetic field. The solid lines are the fitted Gaussian functions. Error bars represent the standard deviation.}
\end{figure}

\begin{figure}[h]
\renewcommand\thefigure{S2}
\includegraphics[width=16cm]{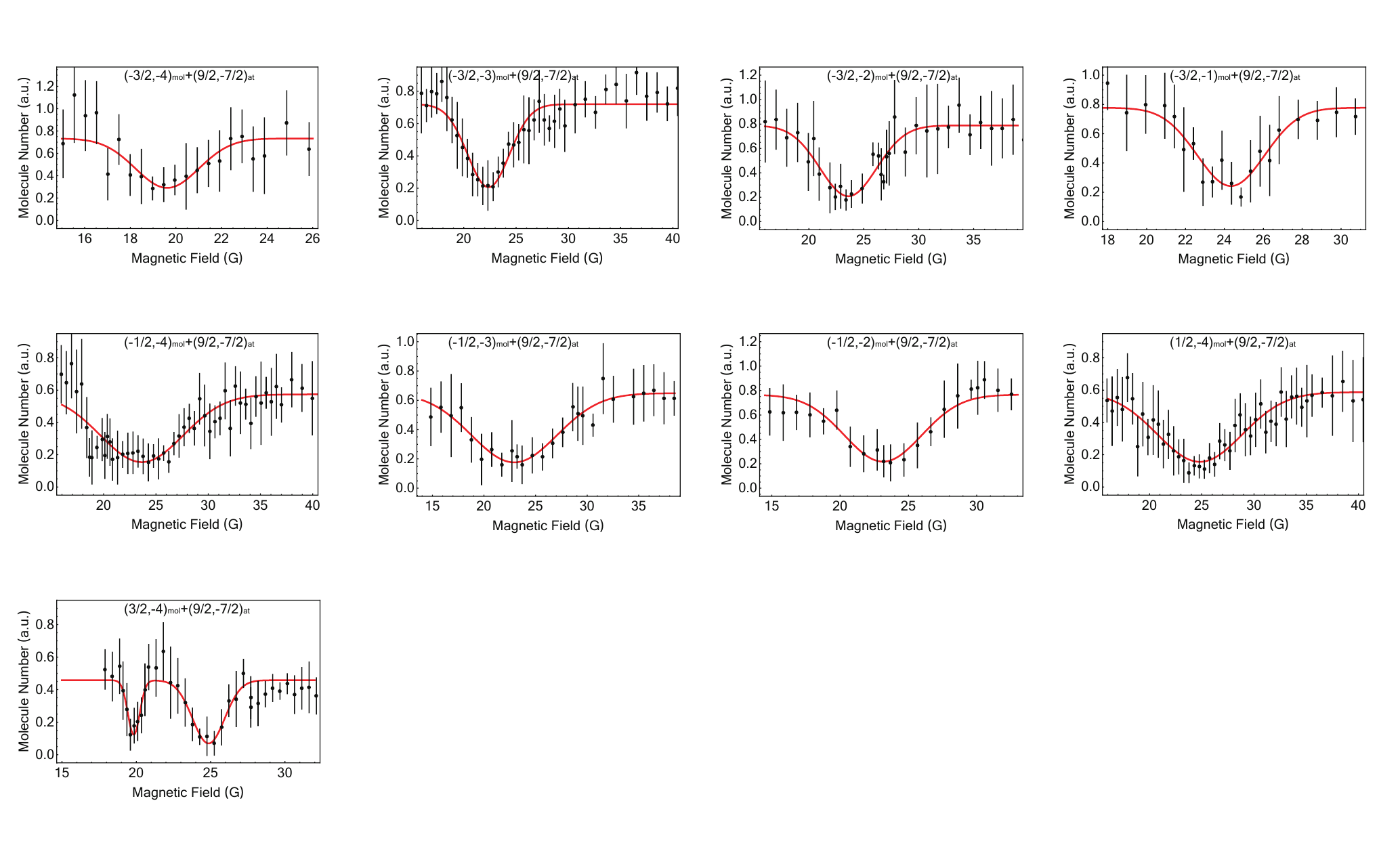}
\caption{\label{fig:sam} The low-field resonances with the initial state of the $^{40}$K atom in $(9/2,-7/2)$. The remaining molecule numbers are shown as a function of the magnetic field. The solid lines are the fitted Gaussian functions. Error bars represent the standard deviation.}
\end{figure}

\begin{figure}[h]
\renewcommand\thefigure{S3}
\includegraphics[width=16cm]{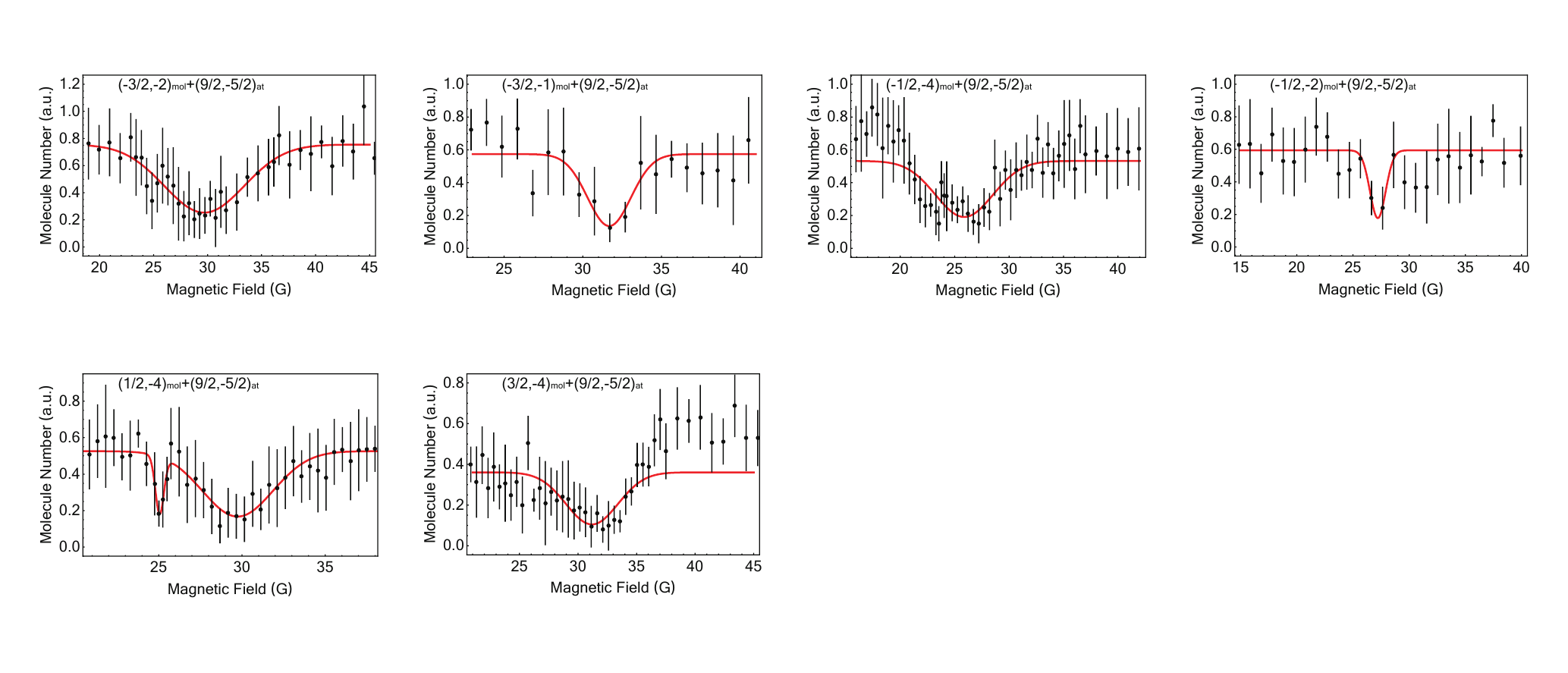}
\caption{\label{fig:sam} The low-field resonances with the initial state of the $^{40}$K atom in $(9/2,-5/2)$. The remaining molecule numbers are shown as a function of the magnetic field. The solid lines are the fitted Gaussian functions. Error bars represent the standard deviation.}
\end{figure}

\begin{figure}[h]
\renewcommand\thefigure{S4}
\includegraphics[width=16cm]{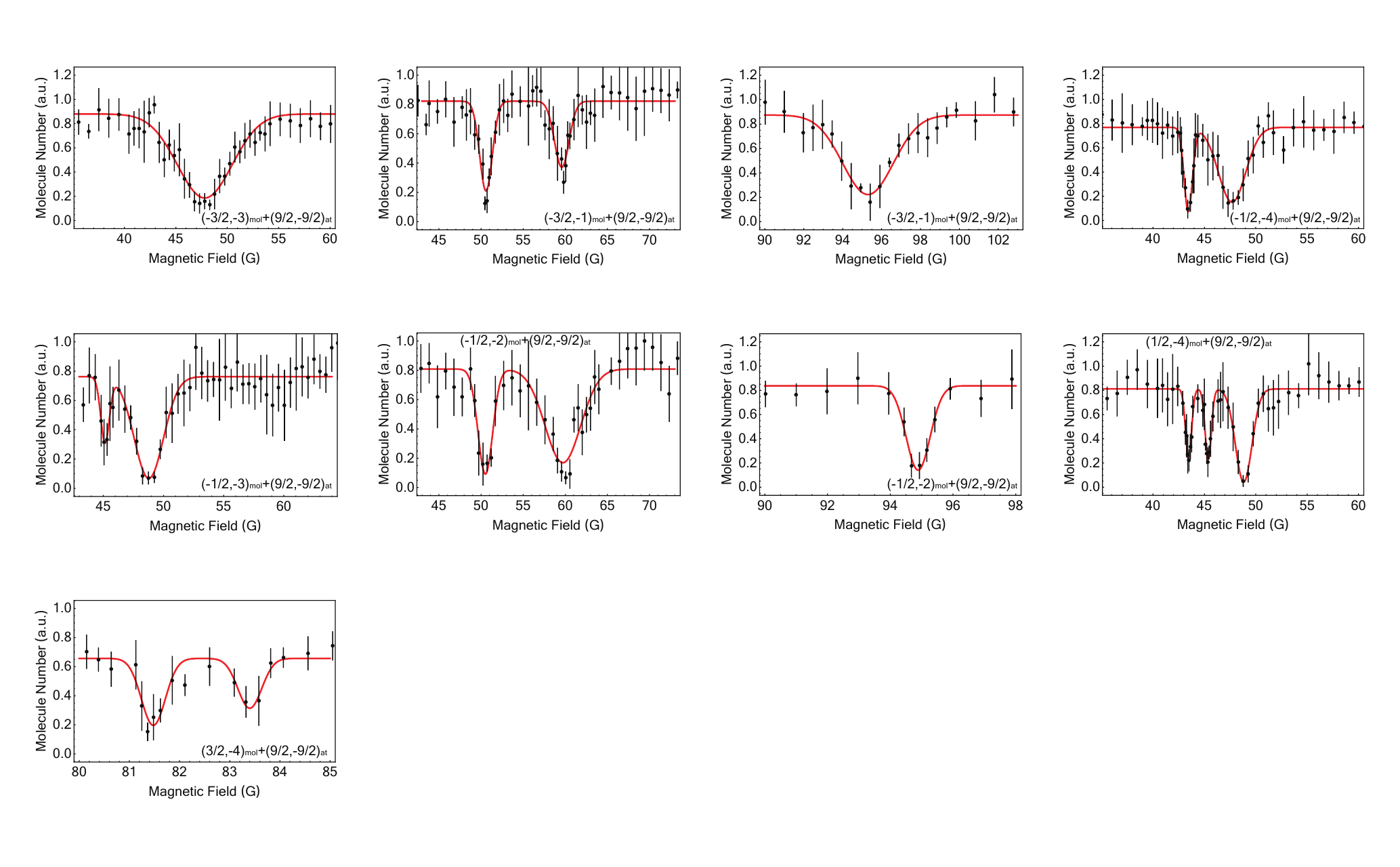}
\caption{\label{fig:sam} The high-field resonances with the initial state of the $^{40}$K atom in $(9/2,-9/2)$. The remaining molecule numbers are shown as a function of the magnetic field. The solid lines are the fitted Gaussian functions. Error bars represent the standard deviation.}
\end{figure}

\begin{figure}[h]
\renewcommand\thefigure{S5}
\includegraphics[width=16cm]{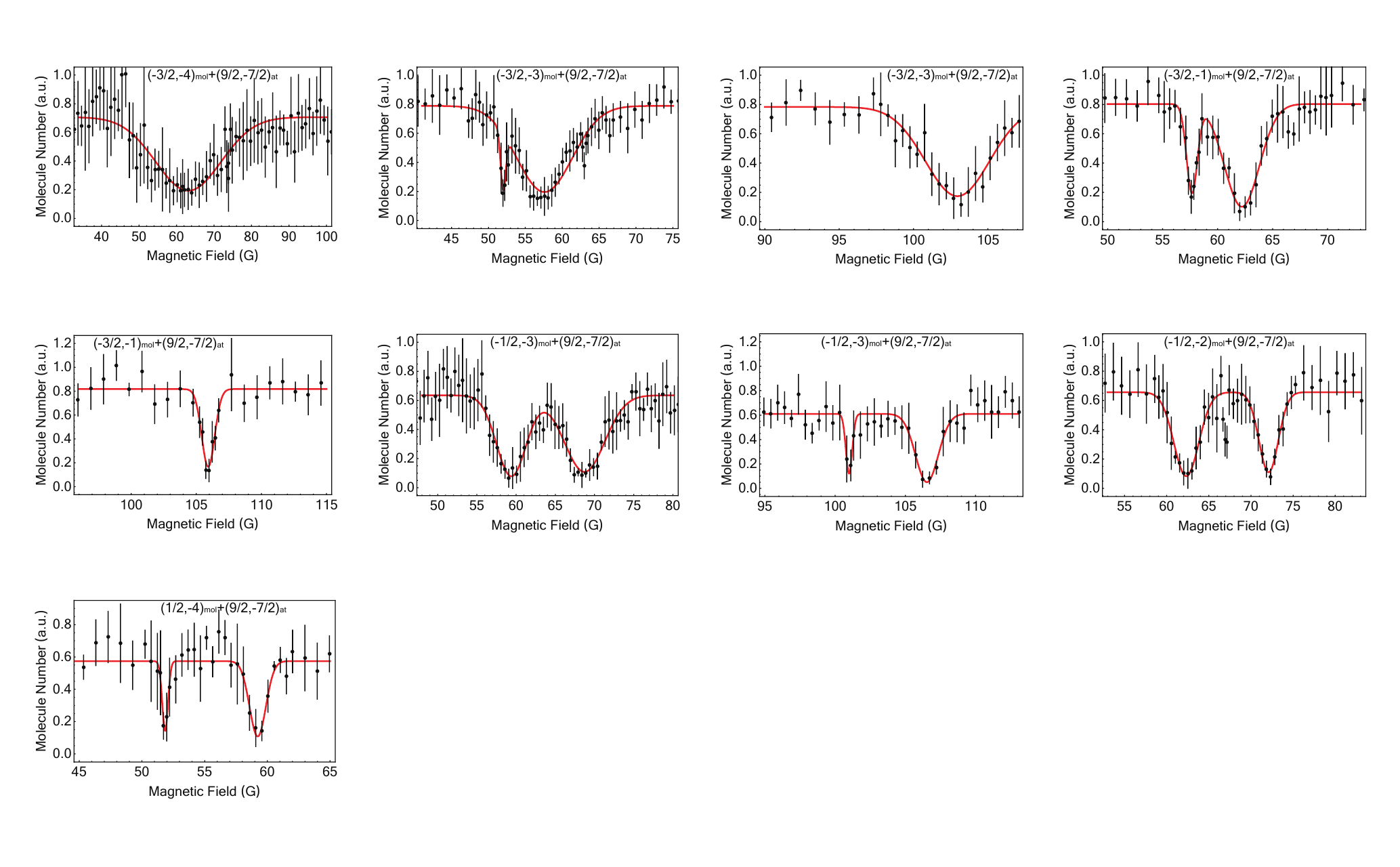}
\caption{\label{fig:sam} The high-field resonances with the initial state of the $^{40}$K atom in $(9/2,-7/2)$. The remaining molecule numbers are shown as a function of the magnetic field. The solid lines are the fitted Gaussian functions. Error bars represent the standard deviation.}
\end{figure}

\begin{figure}[h]
\renewcommand\thefigure{S6}
\includegraphics[width=16cm]{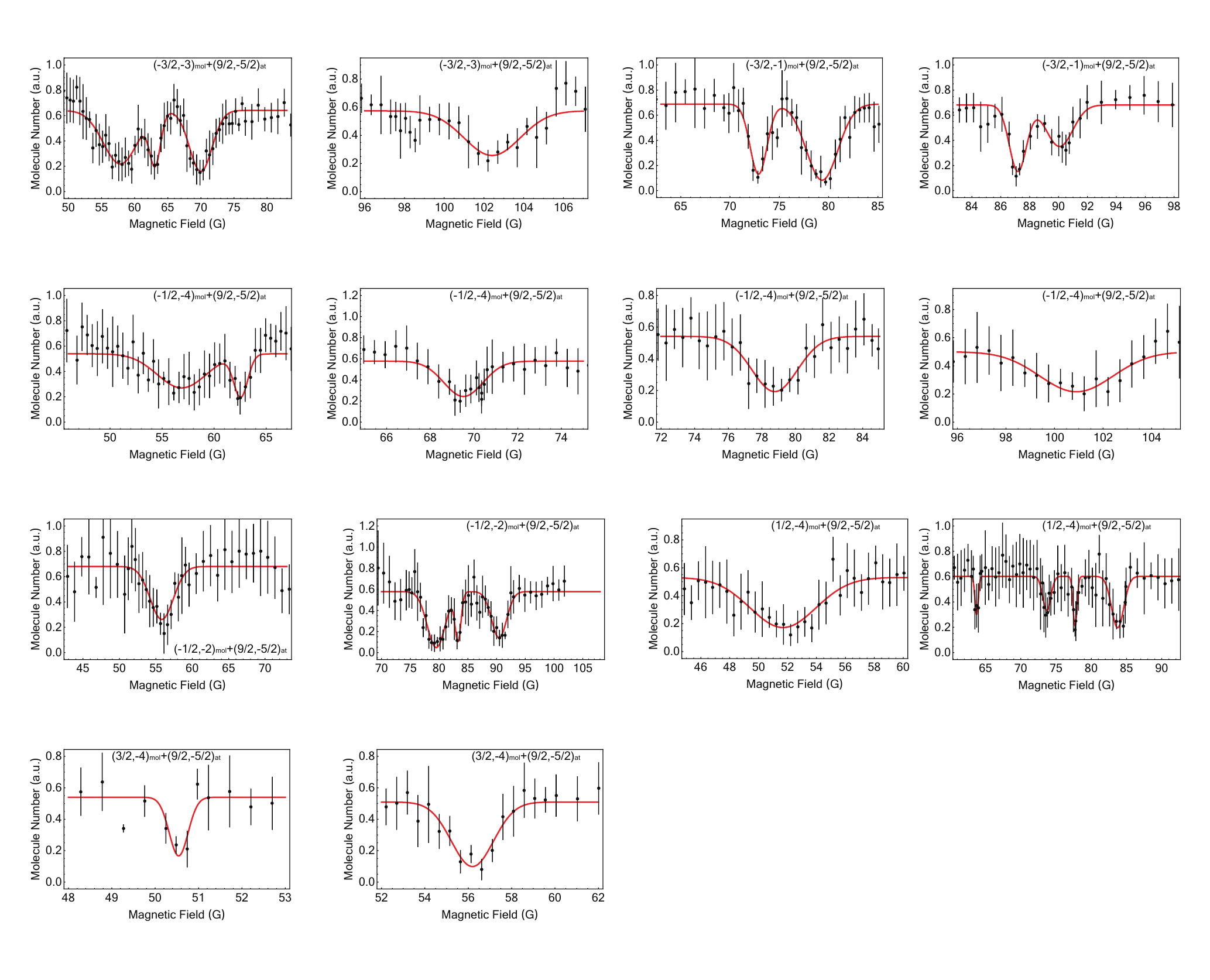}
\caption{\label{fig:sam} The high-field resonances with the initial state of the $^{40}$K atom in $(9/2,-5/2)$. The remaining molecule numbers are shown as a function of the magnetic field. The solid lines are the fitted Gaussian functions. Error bars represent the standard deviation.}
\end{figure}

\end{document}